\documentclass{webofc}
\usepackage[varg]{txfonts}   
\begin{document}
\title{Simultaneous minijets and QGP evolution}
%
%

\author{\firstname{Charles} \lastname{Gale}\inst{1} \and
        \firstname{Sangyong} \lastname{Jeon}\inst{1} \and
        \firstname{Daniel} \lastname{Pablos}\inst{2,3,4} \and
        \firstname{Mayank} \lastname{Singh} \inst{5,6}\fnsep\thanks{\email{mayank.singh@vanderbilt.edu}}
}

\institute{Department of Physics, McGill University,
  3600 University Street, Montreal, QC, H3A 2T8, Canada 
\and
           Departamento de Fisica, Universidad de Oviedo Avda. Federico Garcia Lorca 18, 33007 Oviedo, Spain
\and
            Instituto Universitario de Ciencias y Tecnologias Espaciales de Asturias (ICTEA) Calle de la Independencia 13, 33004 Oviedo, Spain
\and
            INFN, Sezione di Torino, via Pietro Giuria 1, I-10125 Torino, Italy
\and
            Department of Physics and Astronomy, Vanderbilt University, Nashville, TN 37240, USA        
\and
            School of Physics and Astronomy, University of Minnesota, Minneapolis, MN 55455, USA
          }

\abstract{%
  Minijets traversing through the QGP formed in heavy-ion collisions deposit significant amount of energy in the bulk medium. They also create  gradients in temperatures which alter the flow profile and enhance entropy production. We study the effects of minijets in a simultaneous hydro + jet framework and find that inclusion of minijets requires recalibration of transport properties extracted from the model-to-data comparisons.
}
\maketitle
\section{Introduction}
\label{intro}
The standard model of heavy-ion collisions has been very successful in explaining the wide range of observables measured in the heavy-ion collision experiments at the Large Hadron Collider (LHC) and at the Relativistic Heavy-Ion Collider (RHIC) \cite{Gale:2013da}. The essential ingredient of this standard model is relativistic viscous hydrodynamics, which explains the bulk evolution of the deconfined nuclear matter, the quark-gluon plasma (QGP). The properties of QGP are extracted from the heavy-ion collision experiments by doing careful model-to-data comparisons.

For accurate interpretation of experimental data, it is important that all the relevant physics is included in the models. One such essential ingredient, which is often overlooked, is the minijet-medium interaction. Low energy jets (minijets) are created by hard-scatterings in the initial nucleon-nucleon collisions. They traverse through the medium while interacting with it and depositing energy and momentum. The large number of minijets also create large temperature gradients in the system and significantly alter the flow profile while enhancing entropy production \cite{Pablos:2022piv}.

Hydrodynamics is well-suited to describing the low momentum, long wavelength modes of a medium. The minijets effectively act as the higher momentum modes which feed down to hydrodynamic modes due to the non-linear interactions between them. Our approach to dealing with them is to treat minijets as separate particles propagating through the medium losing energy along the way. The lost energy is then treated as hydrodynamized modes and becomes part of the medium via source terms in hydrodynamic equations.

We show the modification of hydrodynamic flow profile as a consequence of minijets. In sec. \ref{sec2}, we describe our simulation framework. We show modifications to hydro in sec. \ref{sec3} and discuss our findings in sec. \ref{sec4}.

\section{Hydro with minijet quenching}
\label{sec2}

We simulate the Pb-Pb collisions at $\sqrt{s_{NN}}= 2.76$ TeV at the LHC. The hard part of our simulation is initialized using the PYTHIA8 framework \cite{Sjostrand:2007gs}. All the hard processes are included in PYTHIA and the spatial location of the collisions is determined by the binary nucleon-nucleon collision positions. The same binary collision positions are utilized to initialize the soft part of the simulation using the IP-Glasma model \cite{Schenke:2012wb}.

The bulk medium is then evolved using the 3+1 D hydrodynamic model  MUSIC \cite{Schenke:2010nt}. The minijets traverse through this medium splitting and losing energy using the strong-weak hybrid model \cite{Casalderrey-Solana:2014bpa}. The splittings happen in the weak coupling limit and the medium modifications to splittings are neglected. The minijet energy loss in medium happens in the strong coupling limit. The energy lost per unit length of QGP traversed is given by \cite{Chesler:2014jva}
\begin{equation}
     \left.\frac{dE}{dx}\right|_{\text{strongly coupled}} = -\frac{4}{\pi}E_{\text{in}}\frac{x^2}{x_{\text{stop}}^2}\frac{1}{\sqrt{x^2_{\text{stop}}-x^2}} \, .
\end{equation}
The $E_{\text{in}}$ is the initial parton energy while $x_{\text{stop}}$ is the distance at which a parton will come to rest in the medium. The stopping distance can be estimated in the strongly coupled limit from holographic calculations \cite{Gubser:2008as,Hatta:2008tx}.

The energy lost by the minijets is smeared by a Gaussian and fed into the medium using a source term. The energy momentum conservation equation essentially modifies to
\begin{equation}
    \partial_{\nu}T^{\mu\nu} = J^{\mu},
\end{equation}
where $T^{\mu\nu}$ is the energy-momentum tensor of the bulk medium and $J^{\mu}$ is the net rate of local energy momentum loss by the minijets.

Hadrons are sampled from the constant temperature hypersurface at 145 MeV using the Cooper-Frye prescription \cite{Cooper:1974mv}. The remaining hard partons are hadronized by the Lund String model in PYTHIA. All the hadrons, soft and hard, are treated together and they undergo decay and cascade within the UrQMD model \cite{Bass:1998ca}.

\section{Modification of transport coefficients}
\label{sec3}


\begin{figure}
\centering
\includegraphics[width=6cm,clip]{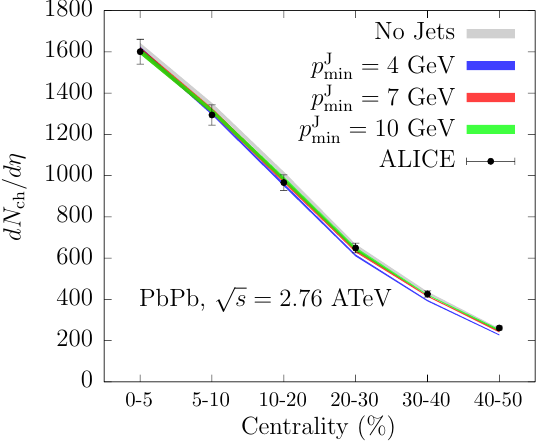}
\includegraphics[width=5.5cm,clip]{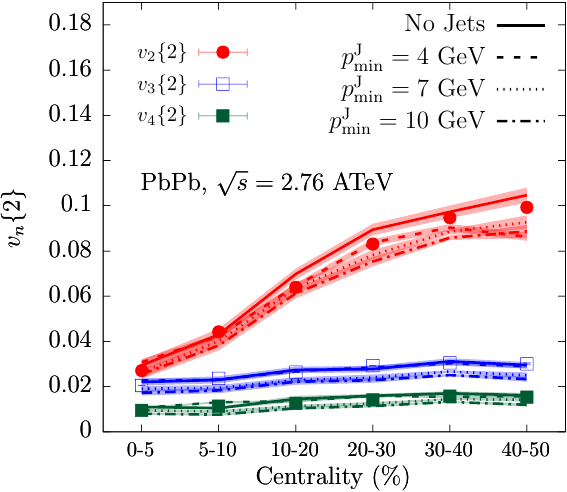}
\caption{Charged hadron multiplicity and harmonic coeficients $v_n$ as a function of centrality. Data from ALICE collaboration \cite{ALICE:2010mlf,ALICE:2011ab}.}
\label{fig-2}       
\end{figure}

There is no clear line separating the hydro and minijets scale. While the soft partons and hard minijets are initialized using distinct physical processes, there is likely rapid interaction and mode mixing at the earliest stages of the collisions making the scale separation ambiguous. So, it becomes hard to motivate the minimum momentum scale at which a parton should be treated as a minijet and not a part of thermal medium. In this study, we treat such a minimum momentum value ($p^{\text{J}}_{\text{min}}$) as a parameter and present out results as a function of $p^{\text{J}}_{\text{min}}$. 

The number of hard scatterings depends on the particular choice of $p^{\text{J}}_{\text{min}}$. The lower the value of $p^{\text{J}}_{\text{min}}$, the larger the number of minijets. They are oriented independent of the background geometry and leave wakes as they pass through a region of QGP. The deposited energy, increased isotropy and increased entropy production from the gradients introduced by wakes necessitate recalibration of model parameters. Here, we just tuned two parameters, the overall energy normalization of the soft initial state ($s_{\text{factor}}$) and the shear viscosity to entropy density ratio ($\eta/s$). Table \ref{tab:newparams} shows the parameters for different $p^{\text{J}}_{\text{min}}$. As more minijets are included, the $s_{\text{factor}}$ must be lowered to account for additional energy and entropy production. As the minijets and their energy deposition tend to increase isotropy, the $\eta/s$ also needs to be reduced to match the data. With the adjustment of just two parameters, simulations with minijets can reproduce the observed data reasonably well (see fig. \ref{fig-2}).

\begin{table}
    \centering
    \begin{tabular}{c|c|c}
      $p^{\text{J}}_{\text{min}}$  & $s_{\text{factor}}$ & $\eta/ s$  \\\hline
        4 GeV & 0.45 & 0.02 \\
        7 GeV & 0.82 & 0.1 \\
        10 GeV & 0.9 & 0.125 \\ 
        No Jets & 0.915 & 0.13 \\
    \end{tabular}
    \caption{Soft initial state normalization and shear viscosity to entropy density for different $p^{\text{J}}_{\text{min}}$.}
    \label{tab:newparams}
\end{table}

\begin{figure}[h]
\centering
\includegraphics[width=12cm,clip]{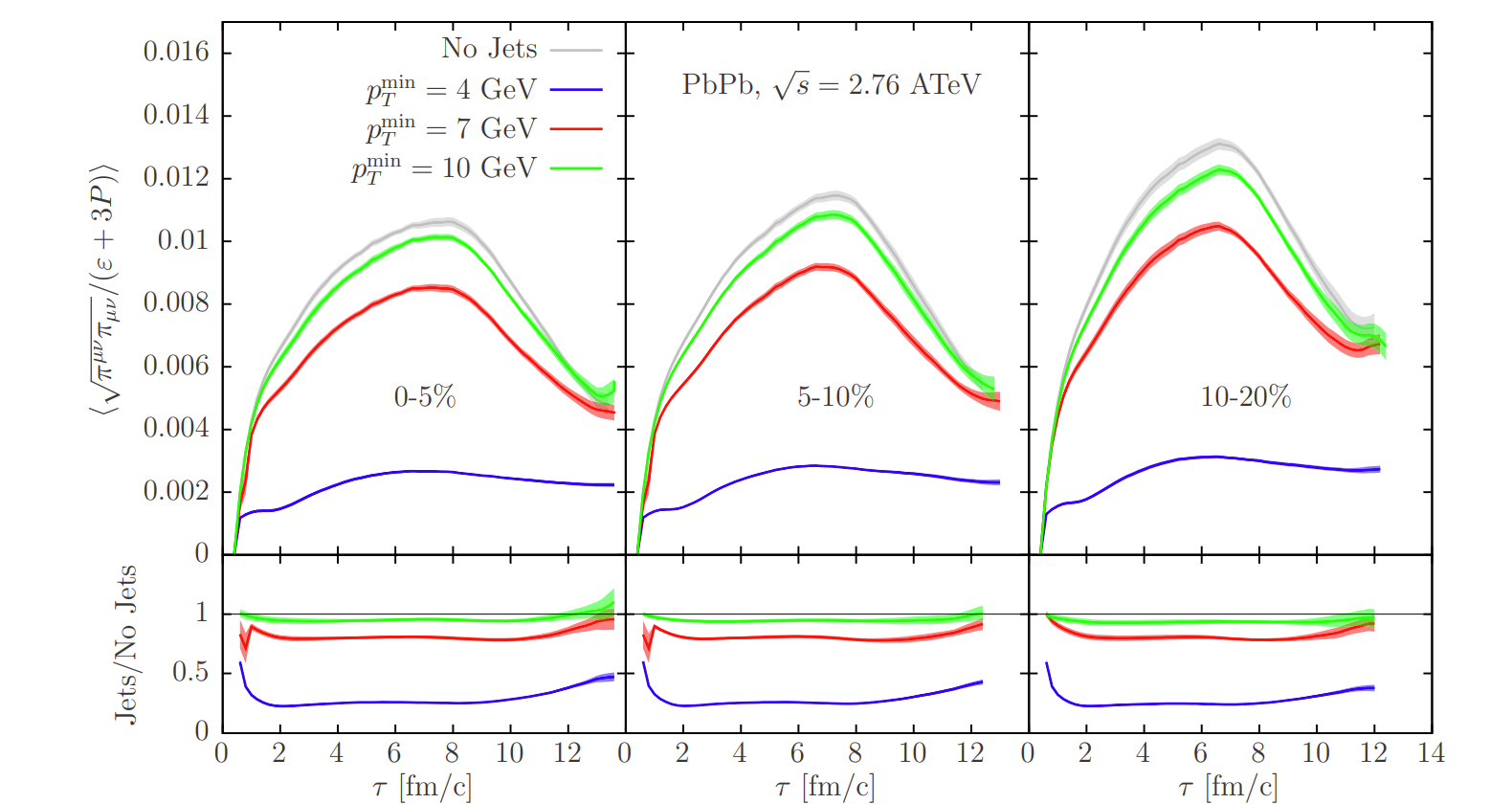}
\caption{Relative size of the shear viscous tensor to the equilibrium energy-momentum tensor as a function of proper time.}
\label{fig-3}       
\end{figure}

Even though the final hadronic profile looks the same, the hydrodynamic profile for systems with different amount of minijets looks significantly different. For example, let us look at the relative magnitude of the shear viscous tensor $\pi^{\mu\nu}$ to the equilibrium energy-momentum tensor (fig. \ref{fig-3}). Including minijets significantly reduces the shear viscous contribution during the hydrodynamic evolution and has an overall effect of bringing the system closer to equilibrium.

\section{Discussion}
\label{sec4}

The minijets interact with the hydrodynamic medium in heavy-ion collisions and they need to be accounted for in simulations. Their inclusion can lead to significant recalibration of model parameters that encode the physical properties of the QCD matter.

With proper tuning of few parameters, models with minijets can explain the observed data reasonably well \cite{Pablos:2022piv}. Even then, the evolutionary history of the system is significantly different. This requires more investigation into observables which can differentiate between the different evolutionary histories. We also need to come up with a gradual soft-hard scale separation model which can simultaneously describe both the soft and hard modes in the intermediate range.

Future model-to-data comparisons will need to account for effects of minijets when extracting physical properties of QGP using heavy-ion collision experiments.



%
%

\vspace{8pt}
\textit{Acknowledgments.} Computations were made on the Beluga supercomputer at McGill University, managed by Calcul Québec and by the Digital Research Alliance of Canada. C.G. and S.J. are supported by the Natural Sciences and Engineering Research Council of Canada under grant numbers SAPIN-2018-00048 and SAPIN-2020-00024 respectively. D.P. has received funding from the European Union’s Horizon 2020 research and innovation program under the Marie Sklodowska-Curie grant agreement No. 754496. M.S. received support from the U.S. DOE under Grant Numbers DE-SC-0024347 and DE-FG02-87ER40328.

\end{document}